\documentclass[aps,prr,reprint,superscriptaddress,longbibliography]{revtex4-1}
\usepackage{amsmath}
\usepackage{graphicx}
\usepackage{amsfonts}
\usepackage{color}
\usepackage{upgreek}
\usepackage{bm}
\usepackage{lipsum}
\usepackage{comment}
\usepackage{enumerate}
\usepackage{dsfont}

\usepackage[table]{xcolor}
\usepackage{textcomp}
\usepackage{amsmath}
\usepackage{amssymb}
\usepackage{graphicx} 
\usepackage{pgf}
\usepackage{epstopdf}
\usepackage{braket}
\usepackage{mathtools}
\usepackage{times}

\usepackage[retainorgcmds]{IEEEtrantools}
\newcommand{\bea}{\begin{eqnarray}}
\newcommand{\eea}{\end{eqnarray}}

\newcommand{\Tr}{\mathrm{Tr}}

\usepackage[colorlinks,linkcolor=blue,urlcolor=blue,citecolor=blue]{hyperref}

\makeatletter
\renewcommand*{\@fnsymbol}[1]{\ensuremath{\ifcase#1\or *\or  
   \mathsection\or \mathparagraph\or \|\or **\or \dagger\dagger
   \or \ddagger\ddagger \else\@ctrerr\fi}}
\makeatother

\begin{document}

\title{Bypassing thermalization timescales in temperature estimation using prethermal probes}

\author{Nicholas Anto-Sztrikacs}
\email{nicholas.antosztrikacs@mail.utoronto.ca}
\affiliation{Department of Physics, 60 Saint George St., University of Toronto, Toronto, Ontario, M5S 1A7, Canada}

\author{Harry J. D. Miller}
\affiliation{Department of Physics and Astronomy, University of Manchester, Oxford Road, Manchester M13 9PL, United Kingdom}

\author{Ahsan Nazir}
\affiliation{Department of Physics and Astronomy, University of Manchester, Oxford Road, Manchester M13 9PL, United Kingdom}

\author{Dvira Segal}
\email{dvira.segal@utoronto.ca}
\affiliation{Department of Chemistry and Centre for Quantum Information and Quantum Control,
University of Toronto, 80 Saint George St., Toronto, Ontario, M5S 3H6, Canada}
\affiliation{Department of Physics, 60 Saint George St., University of Toronto, Toronto, Ontario,  M5S 1A7, Canada}

\date{\today}
\begin{abstract}
We introduce prethermal temperature probes for sensitive, fast and robust temperature estimation.
While equilibrium thermal probes with a manifold of quasidegenerate excited states have been previously recognized for their maximal sensitivity, they suffer from long thermalization timescales. When considering {\it time} as a critical resource in thermometry, it becomes evident that these equilibrium probes fall short of ideal performance.
Here, we propose a different paradigm for thermometry, where
setups originally suggested  for optimal {\it equilibrium} thermometry should  instead be employed as {\it prethermal} probes, by making use of their long-lived quasiequilibrium state. This transient state emerges from the buildup of quantum coherences among quasidegenerate levels. 
For a class of physically-motivated initial conditions, we find that energy measurements of the prethermal state exhibit a similar sensitivity as the equilibrium state. However, they offer the distinct benefit of orders of magnitude reduction in the time required for the estimation protocol. 
Upon introducing a figure-of-merit that accounts for the estimation protocol time, prethermal probes surpass the corresponding equilibrium probes in terms of effective thermal sensitivity, opening avenues for rapid thermometry by harnessing the long-lived prethermal states.     
\end{abstract}
\maketitle

{\it Introduction.--}
The classical notion of thermometry is rooted in the zeroth law of thermodynamics whereby the temperature of a sample is inferred using a small ancillary system, a probe. This probe is brought into contact with the sample until thermal equilibrium is reached \cite{thermometryRev,Thermobook}. Subsequently, the temperature of the sample is inferred by measuring a physically relevant observable of the probe \cite{SensingRev,MetrologyRev}. 
From this protocol, it follows that an ideal probe should offer temperature estimates that are both \textit{accurate} and \textit{rapid}. 
These principles also apply in quantum thermometry, where a classical probe is now replaced by a quantum system. Focusing on precision, the figure of merit quantifying the sensitivity of a probe is captured by the quantum Fisher information, an information-theoretic tool that by the Cramer-Rao inequality upper bounds the signal-to-noise ratio of temperature estimates \cite{thermometryRev,QFI}.

Equilibrium thermometry is a common approach to temperature sensing in the nanoscale regime \cite{Thermometry_book, Dot_probe, SiV_thermo,single-atom-probes}. In this method, the estimation protocol relies on the probe achieving a state of thermal equilibrium with the sample.
These probes are straightforward to operate since energy
measurements maximize their signal-to-noise ratio  \cite{Correa2023,Correa2015}. 
By maximizing the Fisher information of equilibrium probes it was proved
that the most sensitive thermometers are effective two-level systems with a single ground state and $N$ quasidegenerate excited levels \cite{Correa2015,Marti2023}. Furthermore, the effect of multiple probes in both the noninteracting \cite{Correa2017,Correa2023,LandiC19}  and entangled \cite{FermionThermometer,Campbell17,Coh} cases have been analyzed.
However, equilibration times for probes with highly degenerate levels are substantial, and hence,  require a long  time for (thermal) state preparation \cite{Felix_PRE}.
Though time is not a standard parameter in  theoretical thermometry, in practice it plays a crucial role due to the resource nature of thermal state preparation time in quantum sensing \cite{Nemoto,Hayes_2018}. 
To accelerate temperature estimation, {\it transient} temperature estimation schemes, either through pure decoherence processes \cite{Mitchison2020,Mitchison2022,Mitchison2023} or where the probe does not fully thermalize with the sample  \cite{Stace15,Barbieri2018,Barbieri2019,Paternostro,Giovannetti2018,Marti22} have recently been proposed. However, so far such attempts suffer from fundamental issues \cite{Wu2021}. Notably, there is no generic measurement operator of the probe that optimizes the temperature estimate. Furthermore, the Fisher information can strongly fluctuate  as the quantum state of the probe evolves, requiring operating such transient probes in a highly controlled fashion. 

Here, we propose a novel strategy for nanoscale temperature estimation. This approach leverages the accuracy and robustness of equilibrium thermometry with the flexibility and speed of transient thermometry, by putting forward a class of probes that are fast, accurate, robust, and physically motivated. We do this by introducing the concepts of {\it prethermal thermometry} and time-weighted (classical/quantum) Fisher information  as the relevant figure of merit. 
%
Prethermal states materialize when there is a large separation of timescales in the equilibration dynamics \cite{Ueda_2020,Rigol_2019,Gring2012}. In this case, before reaching the true equilibrium state, the system first approaches a {\it quasiequilibrium} state, which is long lived. 
As we demonstrate in this study, such prethermal states offer a robust platform for temperature measurements. 
We consider energy measurements in this work despite their non-optimality due to their simplicity and direct comparison with equilibrium probes demonstrating an advantage despite nonoptimal measurements. 
%
It is possible to repeat the thermometric protocol on the prethermal state numerous times within the extended period that the system requires to achieve full thermalization. 
Repeated measurements leads to a statistical reduction in fluctuations of the temperature estimation, even when the protocol is not finely-tuned for maximizing sensitivity.

Quantum probes exhibiting prethermal states can be engineered in Hamiltonians with nearly degenerate energy levels. In this study, we take the V model and its extension as a case study. The V model comprises three energy levels with a ground state and two nearly degenerate excited states. When weakly coupled to the bath, this system is known to exhibit a long lived transient dynamics with the generation of coherences between excited levels \cite{Tscherbul_2014,Tscherbul_2015,Dodin_2016,Merkli2015,Felix_PRE}. 
This model is particularly relevant since it is an example of an optimal {\it equilibrium} thermal probe \cite{Correa2015}. However, the long equilibration time that this system requires limits its practical utility.
%
%
We analytically compute the quantum Fisher information 
of the  V model and show that for a class of experimentally-motivated initial conditions, the {\it optimal} protocol involves energy measurements of the {\it prethermal} probe. The prethermal state allows for both high sensitivity and a dramatic reduction in the period required to perform the estimation protocol. 
This study thus opens the door to novel architectures for quantum sensors where time as a resource is leveraged using prethermal probes. 


{\it Fundamental limits of quantum thermometry.--}
The objective of thermometry is to acquire information about the temperature of the sample from measurements of the probe. 
To do this, one initially prepares the input state of the probe $\sigma(0)$ and allows it to interact with the sample,  treated as a thermal reservoir, for a duration time $t$. During this interaction, the temperature information is encoded in the probe's state via $\sigma_{\beta}(t) = \mathcal{L}[\sigma(0)]$, where
 $\mathcal{L}$ denotes Markovian time evolution in the superoperator notation \cite{Breuer,Weiss,Strunz20}.
Information about the inverse temperature, the parameter $\beta$, 
is obtained by measuring a physical observable of the probe. 
The sensing precision is given by the Cramer-Rao bound \cite{SensingRev,MetrologyRev}, 
\bea
    \delta \beta \geq [M \mathcal{F}(\beta)]^{-1/2},
    \label{eq:CR}
\eea
where the uncertainty [$\delta \beta = (\langle \beta^2 \rangle - \langle \beta \rangle^2 
)^{-1/2}$] is bounded from below by the inverse 
of the quantum Fisher information (QFI), 
the figure of merit for thermal sensitivity of each of the $M$ independent measurements.
The QFI is obtained from a maximization process over all measurement operators $\hat{O}$ of the classical Fisher information (CFI). It is  useful to define it as $\mathcal{F}(\beta) = \Tr[\hat{L}_{\beta}^2 \sigma_{\beta}(t)]$, where the symmetric logarithmic derivative $\hat{L}_{\beta}$ is given from  $\partial_{\beta} \sigma_{\beta}(t) = \frac{1}{2}\{\hat{L}_{\beta},\sigma_{\beta}(t)\}$ \cite{thermometryRev,SLD}.  


{\it Prethermal thermometry.--}
Prethermal states are stable, long-lived, 
transient states that the system populates before proceeding to its equilibrium state, enabling an {\it expedited} 
temperature estimation at the nanoscale. 
The key advantage of prethermal probes lies in saving the extended time required to prepare a thermal state within the probe.
We do not need to optimize the measurement protocol to see an advantage; 
%
the time savings we get from estimating temperature using prethermal probes allows for many  repetitions of the measurement protocol within the time interval necessary for a single equilibrium measurement. The outcome is the enhancement of the  statistical factor $M$ governing temperature precision, Eq. (\ref{eq:CR}).

Prethermalization effects in unitary quantum systems have been discussed in many studies \cite{Ueda_2020,Rigol_2019,Gring2012,Mori_2018}. 
In open quantum systems, prethermal states develop when there is a large separation of timescales in the relaxation dynamics \cite{De_Vega_2020}. 
In particular, one needs to consider the eigenspectrum of the Liouvillian superoperator responsible for nonunitary dynamics \cite{Felix_PRE}. 
For Markovian evolution, the state of the system can be expanded  
in terms of exponential functions with decay rates captured by the eigenvalues of the Liouvillian $\{\lambda_n\}$.
The timescale to thermalize is dominated by the smallest-magnitude eigenvalue. Prethermal states exist when there is at least one eigenvalue ($\lambda_1$) that  is much smaller in magnitude than the rest of the eigenvalues ($\lambda_{n>1}$),
with the prethermal regime existing between the slowest and second-slowest active modes: $\tau_2 \leq t \leq \tau_1$; 
 $\tau_n = 1/|\lambda_n|$ are the corresponding decay times.

{\it Model.--}
We consider the V model \cite{Dodin_2016,Tscherbul_2014} as a case study for manifesting the advantage of prethermal thermometry over equilibrium thermometry. 
The Hamiltonian consists of the probe, the sample and their coupling term,
\bea
    \hat{H} = \hat{H}_S + \hat{S}\sum_j f_j (\hat{b}_j^{\dagger} + \hat{b}_j) + \sum_j \omega_j \hat{b}_j^{\dagger
    } \hat{b}_j.
    \label{eq:H}
\eea
The probe includes a three level system with two nearly degenerate excited states, $\hat{H}_S = (\nu - \Delta) \ket{2}\bra{2} + \nu\ket{3}\bra{3}$. The energy of the ground state is set to zero.
The sample is composed of independent harmonic oscillators (index $j$) maintained in a thermal state at an inverse temperature $\beta$. The probe couples to the sample via $\hat{S} = \ket{1}\bra{2} + \ket{1}\bra{3} + h.c.$ with interaction energy  $f_j$ captured by the spectral density function $J(\omega) = \sum_j f_j^2 \delta(\omega - \omega_j)$.  We assume that the probe-sample interaction is weak thus non-invasive (contrasting
\cite{Correa2017,Wu2021,Marlon23}). 
As demonstrated in Ref. \cite{Tscherbul_2014}, long lived prethermal states develop when the excited states are nearly degenerate, $\nu \gg \Delta$, with $\Delta$ further being smaller than the thermal relaxation timescale $k$, defined next. 

We adopt the unified Quantum master equation (QME) approach, a simplified Redfield QME that preserves complete positivity and thermodynamic consistency of the quantum dynamics \cite{Anton,Gerry},
and write down equations of motion for the average excited state population [$p(t) = \frac{1}{2}(\sigma_{22}(t) + \sigma_{33}(t))$] and the real and imaginary parts of the coherences  \cite{Felix_PRE}, 
\bea
\dot{p} (t)&=&- k\sigma_{32}^R(t)-\phi p(t)+\frac{\phi-k}{2},
%
\nonumber\\
\dot{\sigma}_{32}^R(t)&=&-k\sigma_{32}^R(t)-\phi p(t)+\Delta\sigma_{32}^I(t)+\frac{\phi-k}{2},
\nonumber\\
\dot{\sigma}_{32}^I(t)&=&-k\sigma_{32}^I(t)-\Delta\sigma_{32}^R(t).
\label{eq:red32I}
\eea
Here, $k=2 J(\nu) [n_B(\nu)+1]$ is the rate constant to transition from the ground state to either of the excited states. The related rate $\phi\equiv k(1+2e^{-\beta \nu})$ is introduced for convenience.  
The Redfield QME builds on the Born-Markov approximation assuming weak (non-invasive) probe-sample coupling and fast dynamics in the sample \cite{Nitzan}. The unified QME further partially secularizes the dynamics,  maintaining coherences only between states close in energy. 
We proceed to solve the unified QME (\ref{eq:red32I}) under the Liouvillian eigenvalues perturbation estimation (LEPE) technique of Ref. \cite{Felix_PRE}. We organize the set of equations as $\frac{d\vec{v}}{dt}= L\vec{v}(t)$, with $\vec{v}(t)=(p(t),\sigma_{32}^R(t),\sigma_{32}^I(t))^{T}$, and 
the Liouvillian constructed from Eq. 
(\ref{eq:red32I}).
Following the LEPE procedure, keeping terms to the lowest order in $\Delta$, the eigenvalues of the Liouvillian are given by 
$\lambda_1\approx\frac{-\phi\Delta^2}{k(k+\phi)}$, $\lambda_2\approx-k$,
$\lambda_3=-(\phi+k)$. 
%
Invoking an exponential ansatz for the Markovian dynamics, 
the solution of the equation of motion for a general initial condition for the population $p(0) = p_0$ and  the real and imaginary coherences, $\sigma_{32}^R(0) = \sigma_{0}^R$, $\sigma_{32}^I(0) = \sigma_{0}^I$, is given by
\bea
\sigma_{32}^R(t)&=&\frac{1}{2(\phi+k)}\Big\{ 
\left[\phi(1 + 2\sigma_{0}^R - 2p_0) -k\right] e^{-\frac{\phi\Delta^2}{k(k+\phi)}t} 
\nonumber\\
&-& \left[\phi(1-2p_0) - k(1+2\sigma_{0}^R)\right] e^{-(\phi+k)t}\Big\},
\label{eq:Csol}
\\
p(t)&=&\frac{\phi-k}{2\phi} 
 \nonumber\\
&-& \frac{1}{2(\phi+k)}\Big\{ 
\frac{k}{\phi}\left[\phi(1 + 2\sigma_{0}^R - 2p_0) -k\right]e^{-\frac{\phi\Delta^2}{k(k+\phi)}t} 
 \nonumber\\
&+& \left[\phi(1-2p_0) - k(1+2\sigma_{0}^R)\right] e^{-(\phi+k)t}\Big\}. 
\label{eq:Psol}
\eea
%
The imaginary part of the coherences is $\mathcal{O}(\Delta)$ and does not contribute significantly to the dynamics, thus to the Fisher information, as we also verify below with simulations. 

\begin{figure}
    \centering
\includegraphics[width=1\columnwidth]{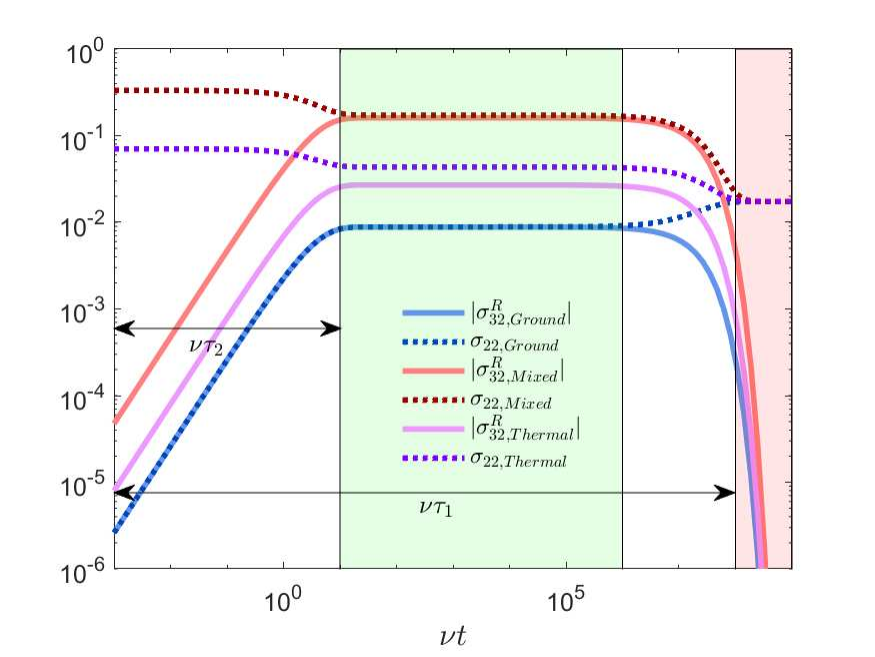}
\caption{The dynamics of coherences (solid) and excited state populations (dotted) in the V model following
Eqs. (\ref{eq:Csol})-(\ref{eq:Psol}) 
displays two dynamical regimes: 
(i) Fast dynamics over $\tau_2$, transitioning the initial condition to the prethermal state. 
(ii) Slow relaxation of the prethermal state extending the period $\tau_1 - \tau_2$ (shaded green region) towards thermal equilibrium (shaded red).
%
The dynamics are studied for three initial conditions: 
(1) $p_0=0$ and $\sigma_0^{R,I}=0$ 
(blue, ``Ground"), 
(2) a maximally mixed state $p_0=1/3$, $\sigma_0^{R,I}=0$ 
(red,``Mix") and (3) a thermal state at ambient temperature $\beta_A$ with $p_0=\frac{e^{-\beta_A \nu}}{1+2e^{-\beta_A \nu}}$ and $\sigma_0^{R,I} = 0$ (purple, ``Thermal"). 
Parameters are $\nu = 1$, $\beta\nu = 4$, $\beta_A\nu = 2.5$,  
$\Delta = 10^{-4}\nu$. We adopt an Ohmic spectral density function,
$J(\omega)=\gamma\omega$, with $\gamma = 0.07$. 
}
\label{fig:Figure1}
\end{figure}
We highlight the appearance of two separate timescales in the dynamics: The short timescales $\tau_{2,3} = \lambda_{2,3}^{-1}$ dictate the time to reach prethermalization.
The long timescale
$\tau_1 = \lambda_1^{-1}$ is associated with full thermalization. The prethermal state (denoted by $\tilde p$ and $\tilde \sigma$) lives during the period 
$\tau_2 \le t \le \tau_1$.
It can be approximated as constant (time independent) in this interval, reminiscent of a true equilibrium state, except with a dependence on the initial conditions, entirely captured by the parameter $\xi\equiv \sigma_{0}^R-p_0$,
\bea
    \tilde{\sigma}_{32}^{R} &=&  \frac{e^{-\beta\nu} + (1+2e^{-\beta\nu})\xi}{2(1+e^{-\beta\nu})},
    \label{eq:CSolPT}
\eea
\bea
    \tilde{p} &=& \frac{e^{-\beta\nu} -\xi}{2(1+e^{-\beta\nu})}.
    \label{eq:PSolPT}
\eea
%
For the density matrix to be physical, the initial conditions are constrained such that $-1\le\xi\le 0$ \cite{supp}.

In Figure \ref{fig:Figure1} we exemplify the dynamics of the V model with three different initial conditions: ground state preparation (blue), maximally mixed state (red) and an ambient (A) thermal state for the probe characterized by an inverse temperature $\beta_A \neq \beta$ (purple). We display the dynamics using the closed-form expressions (\ref{eq:Csol})-(\ref{eq:Psol}). Results were benchmarked (not shown) against numerical results from the Born-Markov Redfield QME and we found a perfect agreement, as expected in the limit of small $\Delta$.
Figure \ref{fig:Figure1} visualizes the emergence of the two separate timescales, $\tau_2$ and $\tau_1$, associated with prethermalization and full thermalization, respectively. 
The prethermal state, captured by Eqs. (\ref{eq:CSolPT})-(\ref{eq:PSolPT})
relies on the buildup of quantum coherences between excited states. It is long lived, lasting for a time interval $(\tau_1 - \tau_2) \approx 10^{6}/\nu$. 
As such, it can serve as a robust alternative to thermal probes, bypassing long thermalization times. Figure \ref{fig:Figure1} further exemplifies  the dependence of prethermal states on initial conditions.
The addition of decoherence  will not affect the lifetime of the prethermal state, 
so long as the decoherence of excited states is correlated, which is expected for atomic or nanoscale probes.

We now compute the CFI and the QFI of a prethermal probe using Eqs. (\ref{eq:CSolPT})-(\ref{eq:PSolPT}) and show that for a class of experimentally-motivated initial conditions, energy measurements are optimal, as in the equilibrium case.
By definition, the CFI is given in terms of the populations $p_j$ of the probe as $\mathcal{I}(\beta) =\sum_j \frac{1}{p_j} (\partial_{\beta} p_j)^2$ \cite{thermometryRev}. We write this measure in terms of the elements of the prethermal density matrix,
\bea
    \mathcal{\tilde I}(\beta) &=& 4\frac{(\partial_{\beta} \tilde{p})^2}{1-2\tilde{p}} + 2\frac{(\partial_{\beta} \tilde{p})^2}{\tilde{p}}.
\eea
As expected, the CFI, which projects onto the eigenbasis of the Hamiltonian, extracts information  from populations only. To obtain information from coherences as well, we need to compute the  QFI,  which by definition is given in terms of the symmetric logarithmic derivative, $\mathcal{\tilde F}(\beta) = \Tr\left[  \hat{L}_{\beta}^2 \tilde{\sigma}_{\beta}\right]$. The prethermal state can easily be diagonalized \cite{supp}.
Importantly, the transformation matrix is independent of temperature, which allows analytic computations of the symmetric logarithmic derivative from which 
the QFI can be computed,
\bea
&&\mathcal{\tilde F}(\beta) = 4\frac{(\partial_{\beta} \tilde{p})^2}{1-2\tilde{p}}
 \nonumber\\
&&+ \frac{2\tilde{p}[(\partial_{\beta} \tilde{p})^2 
+ (\partial_{\beta} \tilde{\sigma}_{32}^{R})^2] - 4\tilde{\sigma}_{32}^{R}(\partial_{\beta} \tilde{p})(\partial_{\beta} \tilde{\sigma}_{32}^{R})  }{\tilde{p}^2 - (\tilde{\sigma}_{32}^{R})^2}.
\eea
This equation was derived while ignoring the imaginary part of coherences, which are $O(\Delta)$ smaller than the real part. Simulations confirm that this assumption is justified.

\begin{figure}
    \centering
\includegraphics[width=\columnwidth]{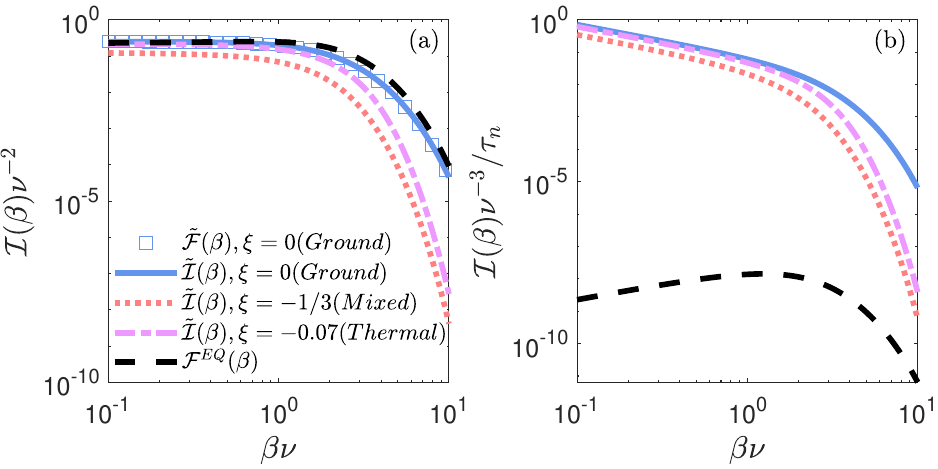}
\caption{Inverse temperature dependence of  the equilibrium and prethermal (classical and quantum) Fisher information.
We display the (a)  Fisher information and (b) the time-weighted Fisher information, $\frac{1}{\tau_2}{\cal \tilde I}(\beta)$ for prethermal thermometry with
%
ground state (blue), maximally mixed state (red) and ambient thermal state (purple) initialization
as well as 
$\frac{1}{\tau_1}{\cal  F}^{EQ}(\beta)$
for equilibrium thermometry (dashed). 
Parameters are the same as in Fig. \ref{fig:Figure1}.}
\label{fig:Figure4}
\end{figure}
Substituting the prethermal state, Eqs. (\ref{eq:CSolPT})-(\ref{eq:PSolPT}) into  the definition of the CFI and the QFI, we get, respectively,
\bea
    \mathcal{\tilde I}(\beta) = \frac{\nu^2e^{\nu\beta}[\xi + 1] }{(1+e^{\nu\beta})^2(1-\xi e^{\nu\beta}) },
    \label{eq:PRE-CFI}
\eea
and
\bea
     \mathcal{\tilde F}(\beta) = \frac{\nu^2e^{\nu\beta}[\xi + 1]}{(1+e^{\nu\beta})^2 }.
     \label{eq:PRE-QFI}
\eea
In the special case where 
$\xi=0$, e.g., a ground state initizalization, the QFI reduces to the CFI.  
The interpretation of this correspondence is that in this case, coherences add no further information about temperature than what populations already bring. We further note that the QFI is strictly greater than the CFI, as required, since $\xi\leq0$ \cite{supp}. 
Optimal QFI  and CFI arise when $\xi = 0$, with the initial and prethermal populations and coherences being equal, resulting in the highest thermal sensitivity of energy measurements.
We contrast the prethermal values of the QFI and CFI with the QFI obtained at equilibrium, where energy measurements are optimal, given by
\bea
\mathcal{F}^{EQ}(\beta) = \frac{2\nu^2e^{\nu\beta}}{(2 + e^{\nu\beta})^2}.
\label{eq:EQ-QFI}
\eea
Comparing this result to Eq. (\ref{eq:PRE-QFI}), we conclude that 
the prethermal probe provides similar precision as the equilibrium one for a single measurement at high temperatures, see also Fig. \ref{fig:Figure4} (a). 
However, the {\it time} required for temperature estimation is substantial in the equilibrium case, given the long thermalization time. 
Therefore, the relevant figure of merit is 
the Fisher information {\it divided by the time it takes to prepare the state over which temperature estimation is performed}. 
This time-weighted quantum Fisher information (TQFI) is given by $ \frac{1}{\tau_2}{\cal \tilde F(\beta)}$ for prethermal thermometry and 
$ \frac{1}{\tau_1}{\cal F}^{EQ}(\beta)$ for equilibrium thermometry. 
Corresponding definitions hold for the time-weighted classical Fisher information (TCFI).
%

We contrast the prethermal CFI and QFI with the equilibrium QFI
in Figure \ref{fig:Figure4} (a).  
The equilibrium QFI (black dashed) is roughly equal to a prethermal QFI with a ground state preparation (blue squares), which
also {\it equals} the CFI (blue solid line). This indicates that energy measurements remain optimal during prethermalization for ground state preparations.
We observe though the loss of sensitivity of energy measurements at low temperature when the V model is initialized
as a maximally mixed state (red) or as an ambient thermal state (purple). 
%
However, as mentioned above, the relevant figure of merit for thermometry should take into account the time it takes to arrive at the state used for metrology. Using the TCFI,
Fig. \ref{fig:Figure4} (b) demonstrates {\it orders of magnitude} advantage in sensitivity for prethermal probes over the corresponding equilibrated probes. 
Most notably, despite energy measurements being suboptimal for non-ground state preparations,
significant time is saved in state preparation. This time can be used to perform  more measurements, offering great reduction in statistical fluctuations through the factor $M$ in Eq. (\ref{eq:CR}).
Namely, for the same period,
the ratio between the number of measurements when using prethermal probes ($\tilde M$) to that number with equilibrated probes ($M^{EQ})$ is $\tilde M/M^{EQ}\approx \tau_1/\tau_2$,
translating to an improvement of $\sqrt{\tau_1/\tau_2}$ in precision. Based on Fig. \ref{fig:Figure1}, $\tau_1/\tau_2\approx 10^6$, thus we achieve a factor of $1000$ 
improvement in thermal sensitivity when using prethermal probes, compared to the equilibrium protocol.

\begin{figure}
    \centering
\includegraphics[width=\columnwidth]{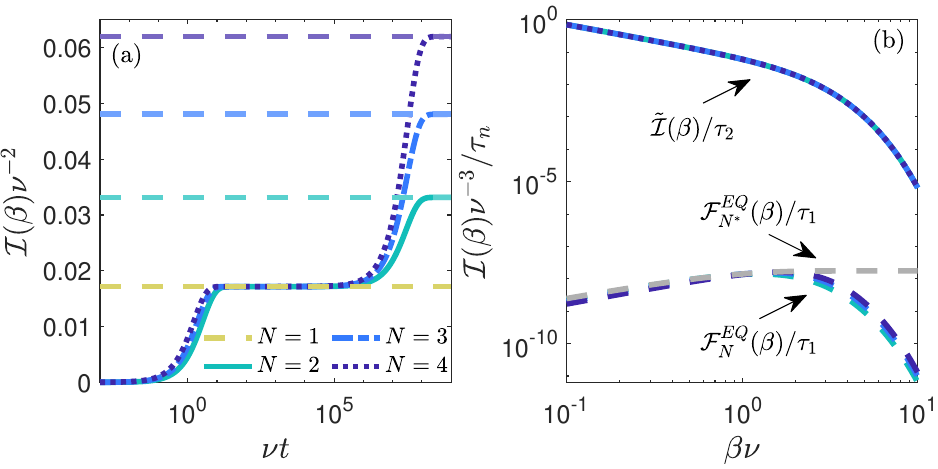}
\caption{$N$ quasidegenerate-level prethermal probes with
 $N=2,3,4$ excited states.
(a) Time dependence of the QFI with a ground state
initialization 
compared to the equilibrium QFI  (constant dashed lines with $N$ growing bottom to top). $N=1$ corresponds to a qubit probe.
(b) Corresponding time-weighted Fisher information as a function of inverse temperature. 
 The grey dashed line is the equilibrium time-weighted Fisher information with $N^*$ excited levels. 
We approximate $\tau_1= \lambda^{-1}_1$ and
$\tau_2=\lambda^{-1}_2$ based on panel (a), assuming these times only mildly change with $N$ for small $N$. Other parameters are the same as in Fig. \ref{fig:Figure1}. 
}
\label{fig:Figure5}
\end{figure}

{\it $N$-level probes.--}
So far, we exemplified the advantage of prethermal thermometry using the V model as a case study. However, our results are more general and robust. We consider the optimal equilibrium probe suggested in  Ref. \cite{Correa2015}, with $N$ nearly-degenerate excited states and a single ground state. We obtain the equilibrium QFI \cite{supp} 
\bea
    \mathcal{F}^{EQ}_N(\beta) = \frac{N\nu^2e^{\nu\beta}}{(N + e^{\nu\beta})^2}.
    \label{eq:FN}
\eea
Notably, this QFI is maximized at $N^*=e^{\beta\nu}$ with 
$\mathcal{F}^{EQ}_{N^*}(\beta)=\nu^2/4$.
Simulating the Redfield QME, we demonstrate in Fig. \ref{fig:Figure5}(a) the time dependence of the QFI for an initial ground state preparation for $N=2$ (teal), $N=3$ (blue), and $N=4$ (purple) excited levels. We find that the long-lived prethermal state is robust, and approximately constant with $N$. 
The equilibrium QFI grows with $N$ in this regime, while the prethermal QFI is constant. 
We further display (gold) the qubit ($N=1$) equilibrium QFI in Fig. \ref{fig:Figure5}(a) showing that the equilibrium qubit probe performs equally well to the prethermal probes. 
We discuss this agreement further in the Supplemental Material \cite{supp}.  
Note that a qubit probe does not display a long-lived transient dynamics. 
%
%
Using the proper measure, which is the TCFI, in Fig. \ref{fig:Figure5}(b),
we find that
${\cal \tilde I}(\beta)/\tau_2$ is about 5 orders of magnitude greater than the equilibrium value Eq. (\ref{eq:FN}), even when using the most optimal equilibrium setup, $ {\cal F}^{EQ}_{N^*}(\beta)/\tau_1$ (grey).
Note that when  $\beta\nu \gtrsim 4$, $N^*>50$, which is impractical. 
This establishes that 
optimal equilibrium thermometers cannot reach the level of precision of transient thermometry when compared with the proper TQFI/TCFI measure.


{\it Conclusions.--}
Systems with a series of degenerate excited states are optimized for sensitivity for equilibrium thermometry, yet they require extremely long times to reach thermal equilibrium due to the development of quantum coherences between degenerate levels. 
Instead, we introduced here the concept of prethermal thermometry, where  long-lived quasi-stationary states  were utilized as robust probes to infer the temperature of a sample.
The dramatic advantage of prethermal probes over their equilibrium counterparts arises from saving extreme long contact times required to reach thermal equilibrium in such setups.
Considering the V model as a case study, we calculated  analytically the QFI and CFI for prethermal probes and demonstrated orders of magnitude improvement in the time-weighted figure of merit for precision.
Generalization to $N$-excited levels demonstrated that the advantage of prethermal thermometry is significant even when compared to an optimized equilibrium probe.
%
Depending on the temperature range, experimental realizations could build on atomic \cite{Vutha18} vibrational  \cite{Molvib} or rotational levels and other engineered or naturally occurring spin impurities \cite{Nir20,NVFabri}. Moreover, such prethermal approaches to thermometry could be leveraged with newly developed Bayesian tools \cite{Correa2021,Correa2022,Marti2022,Landi2021}.
Altogether, prethermal thermometry  drawing upon quantum coherences and other mechanisms offers a novel pathway for achieving accurate, robust and rapid temperature estimation at the nanoscale, particularly suited for weakly-coupled, minimally-invasive scenarios.



{\it Acknowledgments.--} 
D.S. acknowledges support from an NSERC Discovery Grant and the Canada Research Chair program. The work of N.A.S. was supported by Ontario Graduate Scholarship (OGS) and an Alliance Catalyst Quantum Grant. The work of H.J.D.M. has been supported by the Royal Commission for the Exhibition of 1851. 






\bibliography{bibliography} 

\end{document}


\title{Supplemental Material: Bypassing thermalization timescales in temperature estimation using prethermal probes}

\author{Nicholas Anto-Sztrikacs}
\email{nicholas.antosztrikacs@mail.utoronto.ca}
\affiliation{Department of Physics, 60 Saint George St., University of Toronto, Toronto, Ontario,  M5S 1A7, Canada}

\author{Harry J. D. Miller}
\affiliation{Department of Physics and Astronomy, University of Manchester, Oxford Road, Manchester M13 9PL, United
Kingdom}

\author{Ahsan Nazir}
\affiliation{Department of Physics and Astronomy, University of Manchester, Oxford Road, Manchester M13 9PL, United
Kingdom}

\author{Dvira Segal}
\email{dvira.segal@utoronto.ca}
\affiliation{Department of Chemistry and Centre for Quantum Information and Quantum Control,
University of Toronto, 80 Saint George St., Toronto, Ontario, M5S 3H6, Canada}
\affiliation{Department of Physics, 60 Saint George St., University of Toronto, Toronto, Ontario, M5S 1A7, Canada}

\date{\today}
\maketitle

\vspace{5mm}
\renewcommand{\theequation}{S\arabic{equation}}
\renewcommand{\thefigure}{S\arabic{figure}}
\renewcommand{\thesection}{S\arabic{section}}
\setcounter{equation}{0}  
\setcounter{figure}{0}


This Supplemental Material includes the following:
In Sec. \ref{Sec:S1} we provide details on the calculation of the quantum Fisher information (QFI) and the classical Fisher information (CFI) for the V model using prethermal and thermal probes. In Sec. \ref{Sec:S2}, we compare our results to a qubit probe, which does not develop long thermalization times. We discuss in Sec. \ref{Sec:S3} the range of initial conditions for the V model probe, specifically the bounds on the parameter $\xi$. Lastly, in Sec. \ref{Sec:S4} we derive the QFI for an $N$-degenerate level {\it thermal} probe.

\section{Calculation of the classical and quantum Fisher information for a prethermal V model probe}
\label{Sec:S1}

We derive here the QFI and CFI for the  prethermal V probe. Using the near degeneracy of the V model \cite{Tscherbul_2014,Felix_PRE}, the prethermal state can be written in a matrix form in the energy basis of the Hamiltonian as
\bea
\tilde{\sigma}_{\beta} &=&
\begin{bmatrix}
1-2\tilde{p} & 0 & 0 \\
0 & \tilde{p} & \tilde{\sigma}_{32}^{R} \\
0 & \tilde{\sigma}_{32}^{R} &\tilde{p}\\
\end{bmatrix},
\eea
ignoring the small imaginary terms, which are smaller than the real part by a factor $\Delta$. The state
is thus determined by two terms given by
\bea
    \tilde{\sigma}_{32}^{R} 
    &=& \frac{e^{-\beta\nu} + (1+2e^{-\beta\nu})\xi}{2(1+e^{-\beta\nu})},
\eea
and
\bea
    \tilde{p} 
    &=& \frac{e^{-\beta\nu} -\xi}{2(1+e^{-\beta\nu})}.
\eea
In this expression, the parameter $\xi = \sigma_0^R - p_0$ carries information about the initial conditions. As we show in Sec. \ref{Sec:S3}, it satisfies $-1\le\xi\le0$.  $p_0$ is the initial {\it averaged} population of the nearly-degenerate excited levels and $\sigma_0^R$ is the real part of the initial coherences between the excited levels.
From this, the prethermal CFI can be computed from the prethermal populations $\tilde{p}_j$ as 
\bea
\nonumber
    \tilde{I}(\beta) &=& 
    \sum_j \frac{(\partial_{\beta}\tilde{p}_j)^2}{\tilde{p}_j},
    \\ 
    &=& 4\frac{(\partial_{\beta} \tilde{p})^2}{1-2\tilde{p}} + 2\frac{(\partial_{\beta} \tilde{p})^2}{\tilde{p}}.
    \label{eq:SuppCFI}
\eea
To compute the QFI, we make use of the fact that the prethermal state can  be easily diagonalized, written in the form $\tilde{\sigma}_{\beta} = \hat{U}\hat{D}\hat{U}^{\dagger}$ where the matrices $\hat{U}$ and $\hat D$ are given by
\bea
\hat{U} =
\begin{bmatrix}
1 & 0 & 0 \\
0 & -1 & 1 \\
0 & 1 & 1\\
\end{bmatrix}, 
\hat D =
\begin{bmatrix}
1-2\tilde{p} & 0 & 0 \\
0 & \tilde{p} - \tilde{\sigma}_{32}^{R}  & 0 \\
0 & 0 & \tilde{p} + \tilde{\sigma}_{32}^{R}\\
\end{bmatrix}.
\eea
Note that the transformation matrix $\hat{U}$ is independent of the temperature, while the diagonal representation of the state, $\hat D$ contains such a dependency. To compute the QFI, we must first compute the symmetric logarithmic derivative of the prethermal state \cite{SLD}, which is given as the solution to the Lyapunov equation
\bea
    \hat{L}_{\beta} = 2\int_0^{\infty} d\eta e^{-\eta \tilde{\sigma}_{\beta}} (\partial_{\beta} \tilde{\sigma}_{\beta}) e^{-\eta \tilde{\sigma}_{\beta}}.
\eea
Inserting the diagonal representation of the prethermal state into the definition of the symmetric logarithmic derivative and making use of the fact that $\hat{U}$ is temperature independent, this equation can be solved analytically to yield
\bea
\hat{L}_{\beta} = \hat{U} \hat D^{-1} (\partial_{\beta}\hat D)\hat{U}^{\dagger}.
\eea
From this, the QFI for the prethermal state is computed as
\bea
\tilde{\mathcal{F}}(\beta) &=& \text{Tr} \left[ \tilde{\sigma}_{\beta} \hat{L}_{\beta}^2 \right],
\\ \nonumber
&=& \text{Tr} \left[\hat D^{-1}(\partial_{\beta}\hat D)^2 \right],
\\ \nonumber
&=& \sum_j \frac{(\partial_{\beta}d_j)^2}{d_j},
\eea
where $d_j$ are the eigenvalues of the prethermal state, given by the diagonal entries in $\hat D$. This expression can be further expanded to yield an analogous expression to Eq. (\ref{eq:SuppCFI}),
\bea
&&\mathcal{\tilde F}(\beta) = 4\frac{(\partial_{\beta} \tilde{p})^2}{1-2\tilde{p}}
 \nonumber\\
&&+ \frac{2\tilde{p}[(\partial_{\beta} \tilde{p})^2 
+ (\partial_{\beta} \tilde{\sigma}_{32}^{R})^2] - 4\tilde{\sigma}_{32}^{R}(\partial_{\beta} \tilde{p})(\partial_{\beta} \tilde{\sigma}_{32}^{R})  }{\tilde{p}^2 - (\tilde{\sigma}_{32}^{R})^2}.
\eea
It is clear that the QFI equals the CFI when there are no coherences present. More generically,  $\tilde{\mathcal{F}}(\beta) \ge \tilde{\mathcal{I}}(\beta)$.
The derivative terms in the above expressions are given by 
\bea
    \partial_{\beta} \tilde{p} = -\frac{[\xi+1]\nu e^{-\beta \nu} }{2[(1+e^{-\beta \nu})]^2},
\eea
and
\bea
    \partial_{\beta} \tilde{\sigma}_{32}^{R} = -\frac{[\xi+1]\nu e^{-\beta \nu} }{2[(1+e^{-\beta \nu})]^2}. 
\eea
Using these expressions, the prethermal quantum and classical Fisher information can be written in terms of the initial conditions as 
\bea
    \tilde{\mathcal{I}}(\beta) = \frac{\nu^2e^{\nu\beta}[\xi + 1] }{(1+e^{\nu\beta})^2(1-\xi e^{\nu\beta}) },
\eea
and
\bea
     \tilde{\mathcal{F}}(\beta) = \frac{\nu^2e^{\nu\beta}[\xi + 1]}{(1+e^{\nu\beta})^2 }.
     \label{eq:QFI_Pre}
\eea
As mentioned above, the QFI and CFI agree when $\xi=0$.

\section{Qubit thermal probe}
\label{Sec:S2}
In this section, we contrast the prethermal and equilibrium V model probes against a qubit probe. 
First we recall that the qubit probe was generalized to include quasidegenerate excited levels as a strategy to increase sensitivity \cite{Correa2015}. 

Qubit probes do not exhibit long-lived  states. As a result, the qubit probe reaches thermalization in roughly the same timescale that it takes the V model probe to reach its prethermal state. As such, it is instructive to compare their respective quantum Fisher information. 

The qubit probe includes a two-level system with a ground state and a single excited state with total spin splitting $\nu$. The Hamiltonian reads
\bea
    \hat{H} = \frac{\nu}{2}\hat{\sigma}_z + \hat{\sigma}_x\sum_k f_k (\hat{b}_k^{\dagger} + \hat{b}_k) + \sum_k \omega_k \hat{b}_k^{\dagger
    } \hat{b}_k.
\eea
Under the weak-coupling assumption, the qubit thermalizes to a Gibbs state with respect to the system Hamiltonian, such that the symmetric logarithmic derivative can easily be computed as $\hat{L}_{\beta} = \langle \hat{H}_S \rangle_{\beta} - \hat{H}_S$. Furthermore, the QFI is simply the variance of the system Hamiltonian $\mathcal{F}_{TLS}^{EQ}(\beta) = \langle \hat{H}_S^2 \rangle_{\beta} - \langle \hat{H}_S \rangle_{\beta}^2$, where the expectation value is taken with respect to the Gibbs state of the qubit \cite{thermometryRev}. It follows that the equilibrium QFI for the spin-boson model is given by  
\bea
    \mathcal{F}_{TLS}^{EQ}(\beta) =  \frac{\nu^2e^{\nu\beta}}{(1+e^{\nu\beta})^2}. 
\eea
This expression is equal to our prethermal QFI when $\xi=0$, Eq. (\ref{eq:QFI_Pre}), indicating that prethermal probes are comparable to equilibrium ones for a class of initial conditions. 
Overall, the quasistationary nature of the prethermal probes is not a detriment to their performance and is at worst comparable to standard qubit probes. As a result, the prethermal temperature probes offer an alternative to their standard equilibrium counterparts. 




\section{Bound on the initial conditions}
\label{Sec:S3}
In this section, we discuss  bounds on the parameter $\xi = \sigma_0^{R} - p_0$, capturing the dependence of the CFI and the QFI on initial conditions in the V model prethermal state.
Here, $p_0$ is the initial population of the quasi degenerate excited levels and $\sigma_0^{R} $ is the value for the real part of the coherences between the excited levels. 
Constraints on $\xi$ result from the initial density matrix required to fulfill both the positivity and the purity conditions. 
As examples, the lower bound $-1 \le \xi \le 0$ arises, 
 from considering an initial pure state, an equal superposition of the two excited states. The upper bound ($\xi=0$) arises from a ground state preparation.

\begin{figure}
    \centering
\includegraphics[width=\columnwidth]{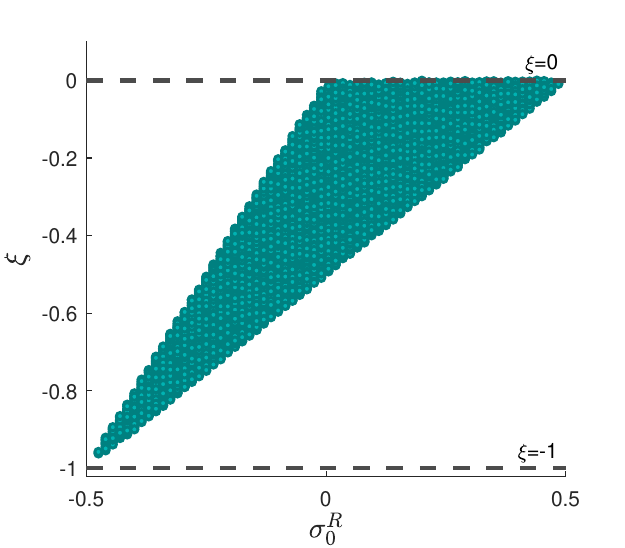}
\caption{
Random samples of density matrices of the form of Eq. (\ref{eq:suppRho_i}) and their associated $\xi = \sigma_0^R - p_0$ values as a function of initial coherences between the excited levels, $\sigma_0^R$. The white region corresponds to initial states where the random sampling produced unphysical states. The (teal) points correspond to physical density matrices. Dashed lines mark the conjectured bounds of $\xi = -1$ and $\xi=0$. We generated $n = 200000$ random sampled with different $\sigma_0^R$ values, with $a,b \in [-1,1]$ and $p_2,p_3 \in [0,1]$.}
\label{fig:Sup}
\end{figure}

We numerically justify the bound by generating and testing a large sample of density matrices that represent possible initial conditions of the form 
\bea
\sigma_0 &=&
\begin{bmatrix}
1-p_2-p_3 & a & b \\
a & p_2 & \sigma_{0}^{R} \\
b & \sigma_{0}^{R} &p_3\\
\end{bmatrix}.
\label{eq:suppRho_i}
\eea
Here, $p_2$ and $p_3$ are the excited state populations and $a$ and $b$ are real parameters quantifying the coherences between the ground and excited states.
For simplicity, the initial populations ($p_2$ and $p_3$) were randomly sampled from a uniform distribution on the interval $[0,1]$ and the coherence parameters $a$ and $b$ were sampled from another uniform distribution on the interval $[-1,1]$.
If the resulting density matrix obeys the purity and positivity constraints, we calculate its $\xi= \sigma_0^{R} - p_0$ value with $p_0=(p_2+p_3)/2$. Results are shown in Figure \ref{fig:Sup}, represented by the teal points. 
We observe that as the initial excited state coherences gets larger in magnitude, the space of physical states shrinks. Our simulations strongly suggest that $-0.5 \le \sigma_0^R \le 0.5$ and that
$\xi$ is lower bounded by -1 and upper bounded by 0, with the bounds depicted by dashed lines. 
We conclude that according to extensive numerics, $-1 \le \xi \le 0$. Proving this result analytically remains a difficult task for a generic initial state.

\section{$N$-degenerate level system: Equilibrium quantum Fisher information}
\label{Sec:S4}
We derive the {\it equilibrium} QFI for an $N$-degenerate level system as presented in the main text and introduced in Ref. \cite{Correa2015}. This setup generalizes the V model.
The QFI can be calculated from the partition function of the system, which consists of a single ground state and $N$ degenerate states at energy $\nu$. It follows that 
\bea
    \mathcal{Z} = 1 + Ne^{-\nu\beta}.
\eea
From this, the QFI can be calculated since the $\mathcal{F}_N^{EQ}(\beta) = \partial_{\beta}^2 \ln \mathcal{Z}$ which implies that 
\bea
    \mathcal{F}_N^{EQ}(\beta) = \frac{\nu^2Ne^{\nu\beta}}{(N+e^{\nu\beta})^2}.
\eea
Using $\beta\nu>1$, this expression predicts that the QFI will grow with $N$ when
$e^{\beta\nu}\gg N$. However, after reaching its maximum sensitivity at $N^*$, for larger $N$ the QFI will decrease monotonically. A maximization calculation shows that for a given temperature, the optimal number of degenerate levels is $N^* = e^{\nu\beta}$. 
The interpretation of this result is that more states are needed at lower temperatures to obtain more  information about precise temperature measurements. For $\beta\nu\approx 1$ $N^*=3$. At lower temperatures, $\beta\nu=4$, one gets $N^*=55$.
 At this value, the quantum Fisher information is given by 
\bea
    \mathcal{F}_{N^*}^{EQ}(\beta)= \frac{\nu^2}{4},
\eea
which is independent of $\beta$. However the figure of merit $\frac{1}{\tau_1}\mathcal
{F}^{EQ}_{N^*}(\beta)$ will still have a nontrivial inverse temperature dependence due to the $\beta$ dependence of the timescales. 

This figure of merit is shown in Fig. 3(b) of the main text using the optimal $N^*$ (dashed line). We find that the times-weighted equilibrium QFI is still
orders of magnitude smaller than the scaled prethermal QFI for experimentally-realizable values of $N$.




\bibliography{bibliography} 
\bibliographystyle{apsrev4-1}
